\documentstyle[epsf,12pt]{article}
 
\newcommand{\AmS}{{\protect\the\textfont2
  A\kern-.1667em\lower.5ex\hbox{M}\kern-.125emS}}

\newcommand{\gsim}{{\protect
  \kern.18em\lower.5ex\hbox{$\stackrel{>}{\sim}$}\kern.25em}}
\newcommand{\lsim}{{\protect
  \kern.17em\lower.5ex\hbox{$\stackrel{<}{\sim}$}\kern.23em}}

\newcommand\be{\begin{equation}}
\newcommand\ee{\end{equation}}
\newcommand\bea{\begin{eqnarray}}
\newcommand\eea{\end{eqnarray}}

\newcommand{\rmO}{\rm O}                                                   

\title{
{\vspace{-0cm} \normalsize
\hfill \parbox{40mm}{CERN 97-239}\\
\hfill \parbox{40mm}{DESY 97-172}}\\[25mm]
The non-perturbative 
${\rmO}(a)$-improved action \\ 
for dynamical Wilson fermions
       \thanks{
               Talk given by K.J. at the International
               Symposium on Lattice Field Theory, 21$-$27 July 1997,
               Edinburgh, Scotland}
        }

\author{Karl Jansen \\ CERN                                  
        1211 Gen\`eve 23 \\ Switzerland \\
\vspace{0.5cm} \\
        Rainer Sommer \\ DESY-IfH \\ Platanenallee 6 \\
        D-15738 Zeuthen \\ Germany                              
    }
\date{~}

\begin{document}

\maketitle

\begin{abstract}
We compute the improvement
coefficient $c_{\rm sw}$ that multiplies the Sheikholeslami-Wohlert
term as a function of the bare gauge coupling for two flavour
QCD.
We discuss several aspects concerning simulations
with improved dynamical Wilson fermions.
\end{abstract}

\vspace{-1cm}

\section{Introduction}

The standard formulation of lattice QCD by Wilson has been used
since the early days of lattice gauge theory. 
However, it is known
that in this formulation the leading discretization errors are
linear in the lattice spacing $a$. Moreover, by testing the
PCAC relation on the lattice, it could be demonstrated that
the effects of these discretization errors 
are most severe and can influence values of  
physical observables strongly \cite{letter}.

In a series of papers \cite{paper1,paper2,paper3} it
was shown that, by implementing Symanzik's improvement programme
\cite{symanzik} for QCD
on-shell and non-perturbatively, one can reach a complete cancellation
of the ${\rmO}(a)$ effects. The advantages of this procedure 
are obvious, and this conference
has seen the improvement programme successfully at work \cite{hartmut}.
The complete improvement programme demands as a first step 
a computation of the
parameter $c_{\rm sw}$ that 
multiplies the Sheikholeslami-Wohlert term \cite{clover}
in the improved action. 
In addition, also 
the coefficients that enter the improved operators 
have to be determined.
By now, in the quenched approximation, a number of these parameters are known as a 
function of the bare
gauge coupling $g_0$ \cite{paper3,Marco,Giulia}. 

In this contribution we want to initiate the computation of the
improvement coefficients for two flavours of
dynamical fermions. 
As a first step we will compute the coefficient $c_{\rm sw}$. 
As is well known, dynamical
fermion simulations are very demanding even with today's computers
and algorithms \cite{karl}. On the other hand, knowing the improvement
coefficients will substantially reduce the computational cost,
since one is allowed to choose larger lattice spacings.

\section{The improvement condition}

The idea of testing the lattice artefacts is to probe
the PCAC relation, which should hold, up to ${\rmO}(a)$ corrections:
\begin{equation} \label{pcac} 
  \partial_{\mu}A_{\mu}^a(x)=2mP^a(x)+{\rmO}(a) \; ,\\
\end{equation}
where $A_{\mu}^a(x)$ denotes the 
isovector axial current and $P^a(x)$ the corresponding
density. 
The quark mass that appears in eq.~(\ref{pcac}) is a
bare current quark mass at scale $1/a$. The important point to notice
here is that the PCAC relation is an operator identity that
can be inserted into arbitrary correlation functions. 

One can make use of this fact to improve the theory:
one tests the PCAC relation
in different correlation functions and demands to obtain
always the same value of $m$.
Using the Schr\"odinger functional,
it was 
demonstrated in ref.~\cite{paper3} how this strategy can be efficiently implemented
to determine $c_{\rm sw}$. Here we follow  ref.~\cite{paper3} closely and 
impose exactly the same improvement condition: 
\begin{equation} \label{condition}
a\Delta M = 0.000277 
\end{equation} 
at $L/a=8$, 
with $\Delta M$ as introduced in ref.~\cite{paper3}. 

The improvement condition eq.~(\ref{condition})
 can in principle be imposed at any (not too large) value
of $M$, where $M$ is a specific definition of the current
quark mass \cite{paper3} derived from eq.~(1).
In order to guarantee a smooth behaviour of
$c_{\rm sw}(g_0)$ one should, however, make a definite choice, where
a natural value is $M=0$. 
Since $\Delta M$ turns out to be a very weak 
function of the quark mass $M$ \cite{paper3,csw_paper}, one may
also compute $\Delta M$ for $|a M| \ll 1$. 
In particular, for the values of $a M$ chosen, here, 
the error introduced 
is negligible compared with the statistical one.

\section{The simulations}

The numerical simulations 
are performed on $16\times  8^3$ lattices, with 
boundary conditions as detailed in \cite{paper3}. 
We use 
the Hybrid Monte Carlo (HMC) algorithm
with the Sexton-Weingarten scheme 
to integrate the classical equations of motion~\cite{SexWei}.
Our implementation of the HMC algorithm                
is described in detail in ref.~\cite{sw_hmc}.
All simulations are performed on the massively parallel Alenia
Quadrics (APE) computers. On the two versions of these machines 
that we have used
with 256 and 512 nodes, we ran 32 and 64 independent
simulations 
in parallel. Combined with a jack-knife method, this allows 
for a realistic error estimate on our observables. 

We have run simulations at eight values of $\beta=6/g_0^2$ in
the range $5.2 \le \beta \le 12.0$. Each simulation
has at least $1280$ molecular dynamics 
trajectories and typically $2500$. 
Keeping the trajectory length fixed to 1, we reach
typical acceptance rates of $95\%$. 
Despite the relatively large acceptance rates, we noticed that
sometimes a system can get stuck and 
does not accept 
a number of (larger than, say, $10$) trajectories. The problem is easily
overcome by performing every $n$ number of trajectories one
with a much smaller step size. Of course, in order to
be able to show that one generates the correct distribution,
the value of $n$ has to be chosen independently of the
Monte Carlo history. We simply kept $n$ fixed in each simulation. 

We applied the improvement condition 
eq.~(\ref{condition}) in the 
small quark mass region, $| a M |<0.01$.\footnote{ 
There is one data 
point where 
$aM=0.023$.} 
With Schr\"odinger functional boundary conditions, 
simulations at
such small quark masses are unproblematic, since the 
massless Dirac operator of the Schr\"odinger functional
with time extent $T$
has a lowest eigenvalue with magnitude
$\lambda_{\rm min} = {\rm const.}/T + {\rmO}(g^2)+ {\rmO}(M)$. 

Owing to the ${\rmO}(g^2)$ terms in $\lambda_{\rm min}$,
the simulations slow down, when $\beta$ is decreased. In detail the reason
for this 
is threefold. First, 
going to smaller 
values of $\beta$ we have to decrease the step size $dt$ 
from, as an example, $dt=0.066$ at $\beta=7.4$
to $dt=0.027$ at $\beta=5.4$. 
Second, 
the condition number $k$ of the preconditioned 
fermion matrix $\hat{Q}^2$ (see e.g. 
ref.~\cite{sw_hmc} for a definition of $\hat{Q}$) 
increases with decreasing $\beta$, as can be seen in fig.~1. 
\vspace{-0mm}
\begin{figure}[t] \label{figure1}
\vspace{-7mm}
\centerline{ \epsfysize=12.8cm
             \epsfxsize=12.8cm
             \epsfbox{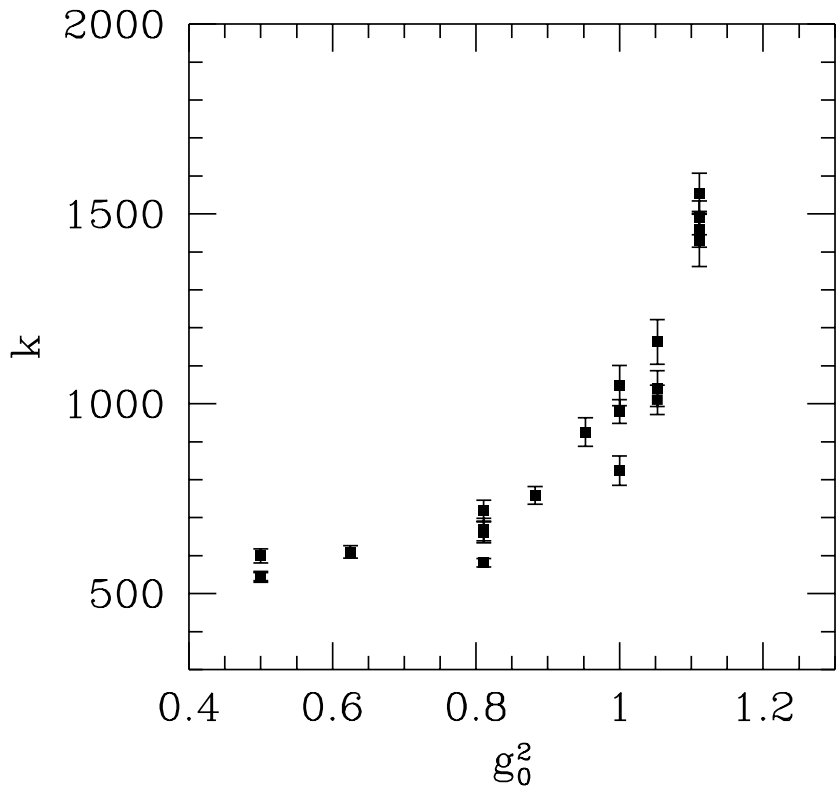}}
\vspace{-14mm}
\caption{The condition number $k$ of the preconditioned 
         fermion matrix $\hat{Q}^2$. Different values 
         of $k$ at the
         same bare gauge coupling $g_0^2$
         correspond to different values of 
         $c_{\rm sw}$.} 
\vspace{-2mm}
\end{figure}
The increasing values of $k$ result 
in a growing number of conjugate gradient
(CG) iterations when going to smaller $\beta$. 
Third, we find an increase of the autocorrelation
time $\tau$ with decreasing $\beta$ for observables such as the lowest 
eigenvalue of $\hat{Q}^2$ or quark correlation functions at a given distance.
Fortunately, it turns out that the autocorrelation time for
$\Delta M$ 
is small,  
$\tau\approx 2$--$4$,
and shows only a weak dependence on $\beta$.

As a rule, we find that the performance of the simulation
algorithm does not significantly depend  
on the value of $c_{\rm sw}$. The only exception
are the autocorrelation times  
$\tau$ for which there are indications that they are particularly large 
when both $c_{\rm sw}$ and $\beta$ are small.

\section{Results}

We determined $\Delta M$ at fixed $\beta$ for various values
of $c_{\rm sw}$. From the --linear-- dependence of $\Delta M$
on $c_{\rm sw}$ we can then extract the slope 
$s=d \Delta M/dc_{\rm sw}$. 
We found 
in practice that the slope is well described by a linear function
of $g_0^2$. In order to extract the desired improvement
coefficients $c_{\rm sw}^{\rm impr}(g_0^i)$ at the eight values of $g_0^i\;, i=1,...,8$, 
where the simulations are performed, we
fit all our data for $\Delta M $ to the form 
\begin{equation}
a\Delta M = s(g_0)\cdot (c_{\rm sw}-c_{\rm sw}^{\rm impr}(g_0^i)) = 0.000277 
\, ,
\end{equation}
where 
\begin{equation} 
s(g_0)=-0.015\cdot(1+s_1 g_0^2) 
\end{equation} 
and $s_1$ as well as $c_{\rm sw}^{\rm impr}(g_0^i)$ are fit parameters.
The results for $c_{\rm sw}^{\rm impr}$ are displayed as the full
symbols in fig.~2. The solid line is a representation of these data, given by 
\begin{equation} \label{pade} 
c_{\rm sw}=  
\frac{1-0.454g_0^2-0.175g_0^4+0.012g_0^6+0.045g_0^8}
              {1-0.720g_0^2} 
\end{equation}

\vspace{-0mm}
\begin{figure}[htb] \label{figure2}
\vspace{-7mm}
\centerline{ \epsfysize=12.8cm
             \epsfxsize=12.8cm
             \epsfbox{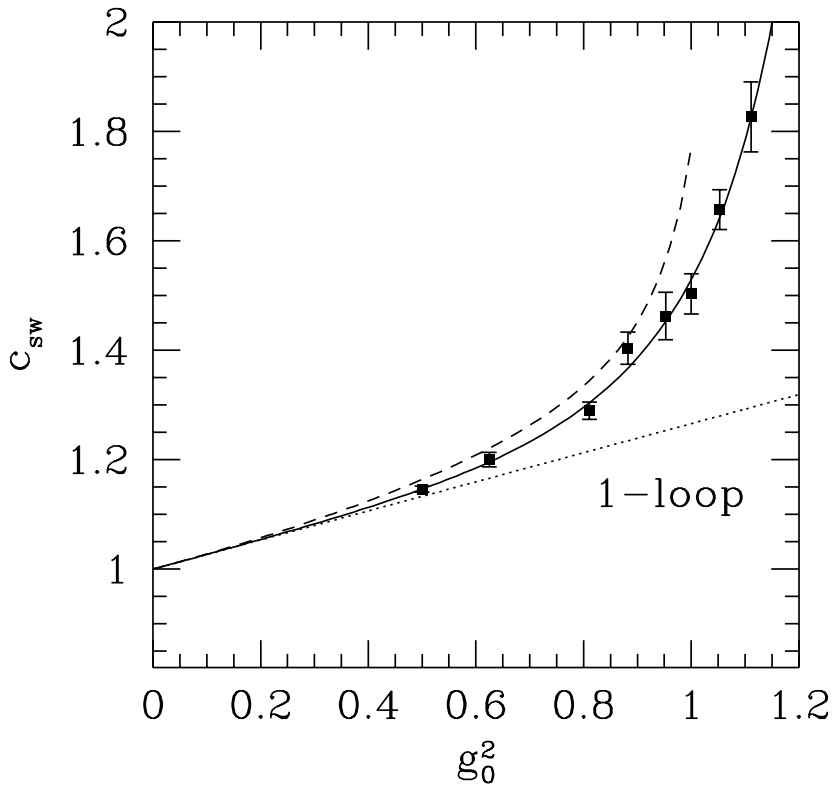}}
\vspace{-14mm}
\caption{The improvement coefficient $c_{\rm sw}$ as a function of
         the bare gauge coupling $g_0^2$. The solid line represents
         eq.~(\ref{pade}). The dotted line is the
         1-loop result \cite{paper2,Wohlert} and the dashed line is the 
         result in
         the quenched approximation \cite{paper3}.} 
\vspace{-0mm}
\end{figure}

As already mentioned, for small values of $\beta$ the simulations
become very costly, and we were not able to perform simulations
at $\beta=5.2$ and small quark masses. We therefore switched to 
the following strategy: we take the parameterization eq.~(\ref{pade}) to
extrapolate a little bit further in $\beta$, to $\beta=5.2$. 
At the 
value of $c_{\rm sw}$ determined in this way, 
we then select a large quark mass, $aM=0.1$ and
try to verify that improvement is at work. 
Indeed, we find for $\beta=5.2$ and 
$c_{\rm sw}=2.02$ that $a\Delta M = -0.0006(9)$. This indicates
that our final result eq.~(\ref{pade}) can safely be used 
for $\beta \geq 5.2$. 
Preliminary studies
of the hadron spectrum in the improved theory
suggest that $\beta \geq 5.2$ yields the range 
of lattice spacings that is of 
interest to computations of 
hadronic properties \cite{Talevi}.

We want to emphasize that although, with our values of 
$c_{\rm sw}$, the ${\rmO}(a)$ terms are cancelled,  
${\rmO}(a^2)$ effects remain and are not negligible
for $\beta \approx 5.2$, as will be
discussed elsewhere \cite{csw_paper}. 

This work is part of the ALPHA collaboration research programme.
We thank DESY for allocating computer time to this project.


\vfill
\begin{thebibliography}{99}                                                     
%
\bibitem{letter} K.~Jansen et.al.,
Phys.Lett.B372 (1996) 275.                     
%
\bibitem{paper1} M.~L\"{u}scher et.al.,   
                 Nucl.Phys.B478 (1996) 365.
\bibitem{paper2} M.~L\"{u}scher and P. Weisz, 
                 Nucl.Phys.B479 (1996) 429.
\bibitem{paper3} M.~L\"{u}scher et.al.,   
                 Nucl.Phys.B491 (1997) 323.
%
\bibitem{symanzik} K.~Symanzik, 
Nucl.Phys.B226(1987)187; 205.
%
\bibitem{hartmut} H.~Wittig, talk at this conference.
%
\bibitem{clover} B. Sheikholeslami and R. Wohlert, Nucl.Phys.B259
 (1985) 572.
%
\bibitem{Marco} M. Guagnelli, talk at this conference.
%
\bibitem{Giulia} G. De Divitiis, talk at this conference.
%
\bibitem{karl} See e.g. 
K.~Jansen, Nucl.Phys.B (Proc. Suppl.) 53 (1997) 127.  
%
%
\bibitem{csw_paper} K.~Jansen and R.~Sommer, in preparation.
%
\bibitem{SexWei}
J.~C.~Sexton and D.~H.~Weingarten, Nucl.Phys.B380 (1992) 665.
%
\bibitem{sw_hmc} K.~Jansen and C.~Liu, 
                 Comput. Phys. Commun. 99 (1997) 221.
%
\bibitem{Wohlert} R. Wohlert, preprint
DESY  87-069 (1987), unpublished
%
\bibitem{Talevi} M. Talevi, talk at this conference.
\end{thebibliography}
\end{document}